\documentclass[
jmp,
reprint,
superscriptaddress,
 amsmath,amssymb,
 aps,
author-numerical,%
]{revtex4-1}

\usepackage{graphicx}
\usepackage{dcolumn}
\usepackage{bm}
\usepackage{xcolor}
\usepackage{bbm}
\usepackage{physics}
\usepackage{appendix}
\usepackage{soul}
\usepackage{epsfig}
\usepackage{float}
\usepackage{multirow}
\usepackage{braket}
\usepackage{lipsum}
\usepackage{hyperref}
\hypersetup{
    colorlinks=true,
    linkcolor=blue,
    filecolor=magenta,      
    urlcolor=blue,
    citecolor = blue,
}
\usepackage[normalem]{ulem}

\begin{document}
\preprint{APS/123-QED}

\renewcommand\st[1]{\ifhmode\unskip\fi}

\title{Complementarity relationship between coherence and path distinguishability in an interferometer based on induced coherence}

\author{Gerard J. Machado}
\email{g.jimenez-machado@imperial.ac.uk}
\affiliation{Department of Physics, Imperial College London, Prince Consort Road, SW7 2AZ London, United Kingdom}
\affiliation{ICFO - Institut de Ciencies Fotoniques, The Barcelona Institute of Science and Technology, 08860 Castelldefels (Barcelona), Spain}%

\author{Lluc Sendra}
\affiliation{ICFO - Institut de Ciencies Fotoniques, The Barcelona Institute of Science and Technology, 08860 Castelldefels (Barcelona), Spain}%

\author{Adam Vall\'es}
\affiliation{ICFO - Institut de Ciencies Fotoniques, The Barcelona Institute of Science and Technology, 08860 Castelldefels (Barcelona), Spain}%

\author{Juan P. Torres}
\email{juanp.torres@icfo.eu}
\affiliation{ICFO - Institut de Ciencies Fotoniques, The Barcelona Institute of Science and Technology, 08860 Castelldefels (Barcelona), Spain}%
\affiliation{Departament of Signal Theory and Communications, Universitat Politecnica de Catalunya, 08034 Barcelona, Spain}%

\date{\today}

\begin{abstract}
We consider an interferometer based on the concept of induced coherence, where two photons that originate in different second-order nonlinear crystals can interfere. We derive a complementarity relationship that links the first-order coherence between the two interfering photons with a parameter that quantifies the distinguishing information regarding the nonlinear crystal where they originated. We show that the derived relationship goes beyond the single-photon regime and is valid for any photon-flux rate generated. We report experimental results in the low photon-flux regime that confirm the validity of the derived complementarity relationship.
\end{abstract}

\maketitle

\section{Introduction}
Quantum interference occurs when there are several alternatives (paths) for an event to happen and there is no way, even in principle, to distinguish between them \cite{mandel1991OL}. This is a fundamental tenet of our understanding of Quantum Theory, and the gold standard for illustrating this idea is the double-slit experiment, in which interference is only observed if the paths are indistinguishable \cite{feynman1964,sudarshan1991}. 

The most common way to quantify quantum interference is the visibility ($\mathcal{V}$) of the fringes of an interference pattern.  However, as we will show here, visibility might not be a good measure of quantum interference, since a loss of visibility ($\mathcal{V}<1$) can also be the result of the interference of mutually coherent optical beams with unequal amplitudes.  To account for any energy unbalance between entities that might interfere, the normalized first-order correlation function $g^{(1)}$ \cite{glauber1963} is a more suitable measure of quantum interference than visibility. 

Distinguishability ($\mathcal{D}$) is associated with the amount of which-way information that allows to link a quantum detection with a particular path. A quantitative formulation of these concepts was investigated half a century ago by Wootters and Zurek \cite{Wootters-Zurek79}, and later by Jaeger \textit{et al.} \cite{Jaeger-etal}. One criterion to assess if a given measure of quantum interference and distinguishability is effective is by verifying its potential inclusion in a complementarity relation of the form $\mathcal{D}^2+\mathcal{V}^2= 1$. Expressions of this type have been derived in various interferometric scenarios \cite{greenberger1988,jaeger1995,englert1996,friberg2017,bagan2016,rempe1998NAT,rempe1998PRL,rempe2000}.

In this paper, our aim is to demonstrate theoretically and experimentally a complementarity relationship in an interferometer based on the idea of induced coherence (see Fig. 1). The concept of induced coherence was introduced in the early 1990s \cite{mandel1991PRL,mandel1991PRA,mandel1992}. When two second-order nonlinear crystals (NLC$_1$ and NLC$_2$) are optically pumped by mutually coherent waves, a pair of signal and idler photons might emerge (signal $s_1$ and idler $i_1$ from NLC$_1$; signal $s_2$ and idler $i_3$ from NLC$_2$) by means of parametric down-conversion (PDC). The idler photon $i_1$ generated in the first crystal experiences losses, which are modeled with a beam splitter with transmissivity $t$ and reflectivity $r$. The signal photons $s_1$ and $s_2$ are made to interfere in a beam splitter and the degree of first-order coherence is the quantity measured. 

Most induced coherence experiments are usually done in the low parametric gain regime (weak pumping) so that paired photons are expected to be generated only in either of the two crystals. Induced coherence can be observed in the high parametric gain regime of down-conversion \cite{belinsky1992,wiseman2000}, when signal and idler pairs are generated in both crystals simultaneously. There is an on-going  discussion about the quantum character of induced coherence in the low gain regime when compared with the high gain regime \cite{shapiro2015,zeilinger2019}. In this paper, we put forward a measure of distinguishability $\mathcal{D}$ that is valid in both parametric gain regimes, which is based solely on second-order correlation functions $g_{si}^{(2)}$ between signal and idler photons. This result reveals that the concept of distinguishability, typically considered applicable only in the single-photon regime, can also be broadened to the high parametric gain regime.

\section{Description of induced coherence in the Schr\"odinger picture}
In the low parametric gain regime of PDC, one can safely assume that paired photons are generated in one nonlinear crystal or in the other, but never simultaneously in both.  The quantum state of signal-idler photons at the output before measurement (see Fig. 1) is
\begin{eqnarray}
    & & |\Psi \rangle_{si}=\frac{1}{\sqrt{2}}\,\ket{1}_{s_1} \ket{0}_{s_2} \Big[ r \ket{1}_{i_2}\ket{0}_{i_3}+t \ket{0}_{i_2}\ket{1}_{i_3} \Big]  \nonumber \\
    & & +\frac{1}{\sqrt{2}}\,\ket{0}_{s_1}\ket{1}_{s_2}  \ket{0}_{i_2}\ket{1}_{i_3}.
\end{eqnarray}
The first term corresponds to the generation of signal-idler photon pairs in the first nonlinear crystal, and the second term corresponds to the generation in the second nonlinear crystal. The quantum state that describes the idler photon if the signal-idler pair is generated in the first nonlinear crystal, in the basis $\{\ket{1}_{i_2}\ket{0}_{i_3}, \ket{0}_{i_2}\ket{1}_{i_3}\}$, is
\begin{equation}
    |\Psi_1 \rangle=r|1 \rangle_{i_2}|0 \rangle_{i_3}+t|0 \rangle_{i_2}|1 \rangle_{i_3}.
\end{equation}
If the signal-idler pair is generated in the second nonlinear crystal, the quantum state of the idler photon is 
\begin{equation}
    |\Psi_{2} \rangle= |0 \rangle_{i_2}|1 \rangle_{i_3}.
\end{equation}
The distinguishability $\mathcal{D}$ between the two events can be quantified as the trace distance \cite{englert1996} $\mathcal{D}=\sqrt{1-|\langle \Psi_1|\Psi_2\rangle|^2}$, so the distinguishability $\mathcal{D}$ is
\begin{equation}
    \mathcal{D}=|r|=\sqrt{1-|t|^{2}}.
\label{eq:disting}
\end{equation}
The interference between the signal photons ($s_1$ and $s_2$) generated in the two different nonlinear crystals (see Fig. 1) is measured by recombining them in a beam splitter. The visibility of the interference fringes as a function of the delay is the quantity measured. In the low parametric gain regime, the visibility is $\mathcal{V}=|t|$ \cite{mandel1991PRL,mandel1992,lemos2014,adam2018}. Thus, distinguishability and visibility are related through the equality $\mathcal{D}^{2}+\mathcal{V}^{2}=(1-|t|^{2})+|t|^{2}=1$.

 \begin{figure}[t!]
 \includegraphics[width=\linewidth]{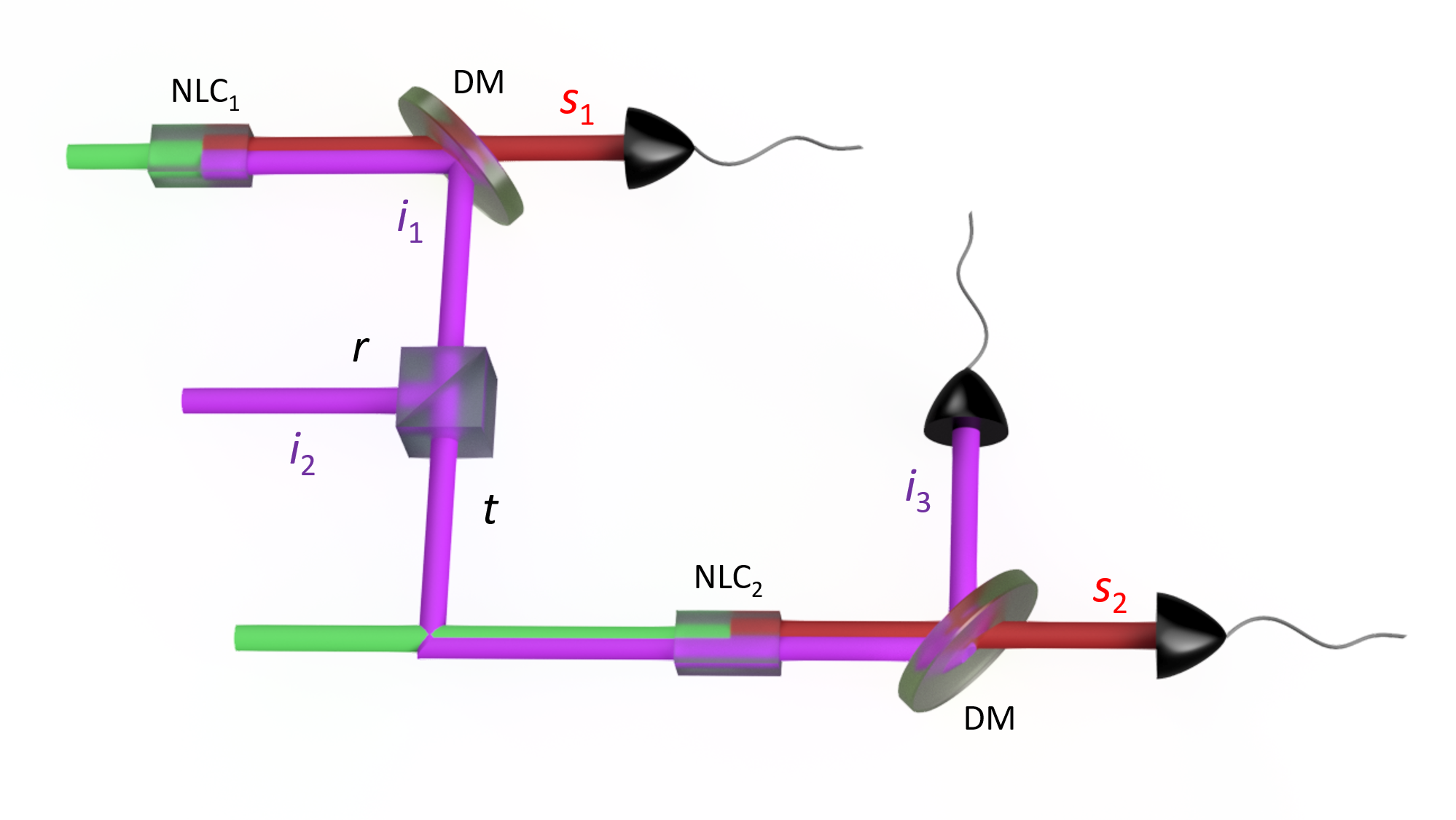}
 \caption{\label{fig:dist_scheme} Sketch of an induced coherence setup. NLC$_{1,2}$: nonlinear crystals; $DM$: dichroic mirrors; $i_2$: idler photon $i_1$ reflected by the BS; $i_3$: idler photon generated in NLC$_2$, or generated in NLC$_1$ and transmitted at the BS. Signal photons $s_1$ and $s_2$ can be made to interfere by recombining them in a beam splitter. The change of the power measured at one output port of the beam splitter as function of the length difference of the paths taken by photons $s_1$ and $s_2$ is a measure of the degree of first-order coherence between the photons.}
 \end{figure}

\section{Description of induced coherence in the Heisenberg picture}
In the process of parametric down-conversion, the relationships between the input signal and idler annihilation operators, $\hat{b}_{s}$ and $\hat{b}_{i}$, and the output operators, $\hat{a}_{s}$ and $\hat{a}_{i}$, are described (in the single-mode approximation) by the Bogoliubov transformations \cite{navez2001,brambilla2004,kolobov2007,torres2011}
\begin{eqnarray}
& & \hat{a}_{s}=U_{s}\,\hat{b}_{s}+V_{s}\,\exp(i \varphi_p)\,\hat{b}_{i}^{\dagger}, \nonumber \\
& & \hat{a}_{i}=U_{i}\,\hat{b}_{i}+V_{i}\,\exp(i \varphi_p)\, \hat{b}_{s}^{\dagger},
\end{eqnarray}
where
\begin{eqnarray}
& & U_{s,i}=\cosh (\sigma L)\,\exp(ik_{s,i} L), \hspace{1cm}  \nonumber \\
& &  V_{s,i}=-i\sinh (\sigma L)\,\exp(ik_{s,i} L), 
\end{eqnarray}
$k_{s,i}$ are wavenumbers for the signal and idler waves, $\varphi_p$ is the phase of the pump beam, $L$ is the length of the nonlinear crystal, and $G=\sigma L$ is the parametric gain \cite{machado2020, chekhova2016}. Notice that $|U_{s,i}|^2-|V_{s,i}|^2=1$.

The idler $i_{1}$ transforms as $\hat{a}_{i_1} \Longrightarrow t\hat{a}_{i_1}+\hat{f}$ due to its interaction with a lossy object.  The operator $\hat{f}$ fulfills $[\hat{f},\hat{f}^{\dagger}]=1-|t|^{2}$ \cite{haus2000,boyd2008OC}. After applying the Bogoliubov relationships twice and taking into account the presence of the beam splitter in the idler $i_1$ path,  we obtain that the output operators are
\begin{eqnarray}
    & & \hat{a}_{s_1}=\Big\{ U_{s}\hat{b}_{s}+V_{s}\exp( i\varphi_{p_1})\hat{b}_{i}^{\dagger} \Big\} \exp(i \varphi_{s_1}),  \nonumber \\  
    & & \hat{a}_{s_2}=\Big\{ U_{s}\hat{c}_{s}+t^* V_i^* V_s\exp( i\varphi_{p_2}-i\varphi_{p_1}-i\varphi_{i_1})\hat{b}_{s} \nonumber \\
    & & +t^* U_{i}^* V_{s}\exp( i\varphi_{p_2}-i\varphi_{i_1})\hat{b}_{i}^{\dagger} \nonumber \\
    & & +V_{s}\exp(i \varphi_{p_2})\hat{f}^{\dagger}\Big\}\exp(i \varphi_{s_2}), \nonumber \\
    & &  \hat{a}_{i_3}=\Big\{ t U_{i}^2\exp(i \varphi_{i_1})\hat{b}_{i}+V_{i}\exp( i\varphi_{p_2}) \hat{c}_{s}^{\dagger}+U_{i}\hat{f}  \nonumber \\
    & &+t U_{i} V_{i}  \exp( i\varphi_{p_1}+i\varphi_{i_1})\hat{b}_{s}^{\dagger}  \Big\}\exp(i\varphi_{i_3}). 
    \label{operators}
\end{eqnarray}
$\varphi_{p_1}$ and $\varphi_{p_2}$ are the phases of the pump beam at the first and the second nonlinear crystals, respectively. $\varphi_{s_1}$, $\varphi_{s_2}$, $\varphi_{i_1}$ and $\varphi_{i_3}$ are phases acquired by signals $s_1$ and $s_2$, and idlers $i_1$ and $i_3$, during propagation. 

The flux rates of the signal and idler photons generated are $\langle \hat{a}_{s_1}^{\dagger}\hat{a}_{s_1} \rangle = |V|^{2}$, $\langle \hat{a}_{s_2}^{\dagger}\hat{a}_{s_2} \rangle = |V|^{2}\big( 1+ |t|^{2}|V|^{2} \big)$ and $\langle \hat{a}_{i_3}^{\dagger}\hat{a}_{i_3} \rangle = |V|^{2}\big( 1+ |t|^{2}|U|^{2} \big)$, where
$|U|\equiv |U_s|=|U_i|$ and $|V|\equiv |V_s|=|V_i|$. We calculate the normalized second-order correlation functions $g_{mn}^{(2)}$ ($m=1,2$ and $n=3$) as
\begin{equation}
 g_{mn}^{(2)}=\frac{\langle \hat{a}_{m}^{\dagger} \hat{a}_{n}^{\dagger} \hat{a}_{n} \hat{a}_{m} \rangle}{\langle \hat{a}_{m}^{\dagger} \hat{a}_{m} \rangle\, \langle \hat{a}_{n}^{\dagger}\hat{a}_{n}\rangle},
\label{g13def}
\end{equation}
between signal $s_{1}$ and idler $i_3$ [$g_{13}^{(2)}$], and signal $s_{2}$ and idler $i_3$ [$g_{23}^{(2)}$]. After a lengthy but otherwise straightforward calculation, one can show that
\begin{equation}
 g_{13}^{(2)}=1+\frac{|t|^2\,|U|^4}{1+|t|^2 |U|^2}\, \frac{1}{|V|^2},
 \label{eq:g13}
\end{equation}
and
\begin{eqnarray}
    & & g_{23}^{(2)}=1+ \label{eq:g23} \\
    & &+\frac{|t|^{4}|U|^{6}+2|t|^{2}|U|^{4}-2|t|^{4}|U|^{4}+|U|^{2}(1-|t|^{2})^{2}}{|V|^{2} \big[ 1+ |t|^{2}|V|^{2} \big]\big[ 1+|t|^{2}|U|^{2}\big]}.  \nonumber
\end{eqnarray}
In the low parametric gain regime ($G \ll 1$), we can approximate $|U|^2 \sim 1$ and $|V|^4 \ll |V|^2$, so that
\begin{equation}
 g_{13}^{(2)}=1+\frac{|t|^2}{1+|t|^2}\, \frac{1}{|V|^2},
 \label{g13low}
\end{equation}
and
\begin{eqnarray}
    & & g_{23}^{(2)}=1+\frac{1}{1+|t|^2}\, \frac{1}{|V|^2}.
\label{g23low}
\end{eqnarray}
If we compare Eq. (\ref{eq:disting}) ---the value of the distinguishability $\mathcal{D}$ obtained in the Schr\"odinger picture in the low parametric gain regime--- with Eqs. (\ref{g13low}) and (\ref{g23low}), also valid in the low parametric gain regime, one can easily verify that the distinguishability $\mathcal{D}$ can be written as
\begin{equation}
    \mathcal{D}=\sqrt{\frac{g_{23}^{(2)}-g_{13}^{(2)}}{g_{23}^{(2)}-1}}.
\label{eq:distinguishability}
\end{equation}
Equation (\ref{eq:distinguishability}) is the first important result of this paper. It relates the distinguishability $\mathcal{D}$ to the measurement of the second-order correlation functions between $s_{1}$ and $s_{2}$, and the idler, which acts as a which-path detector. Equation (\ref{eq:distinguishability}) is an experimentally measurable quantity of the distinguishability between the path taken by signals $s_1$ and $s_2$, after they are combined with the help of a beam splitter. 

 \begin{figure}[t!]
 \centering
 \includegraphics[width=\linewidth]{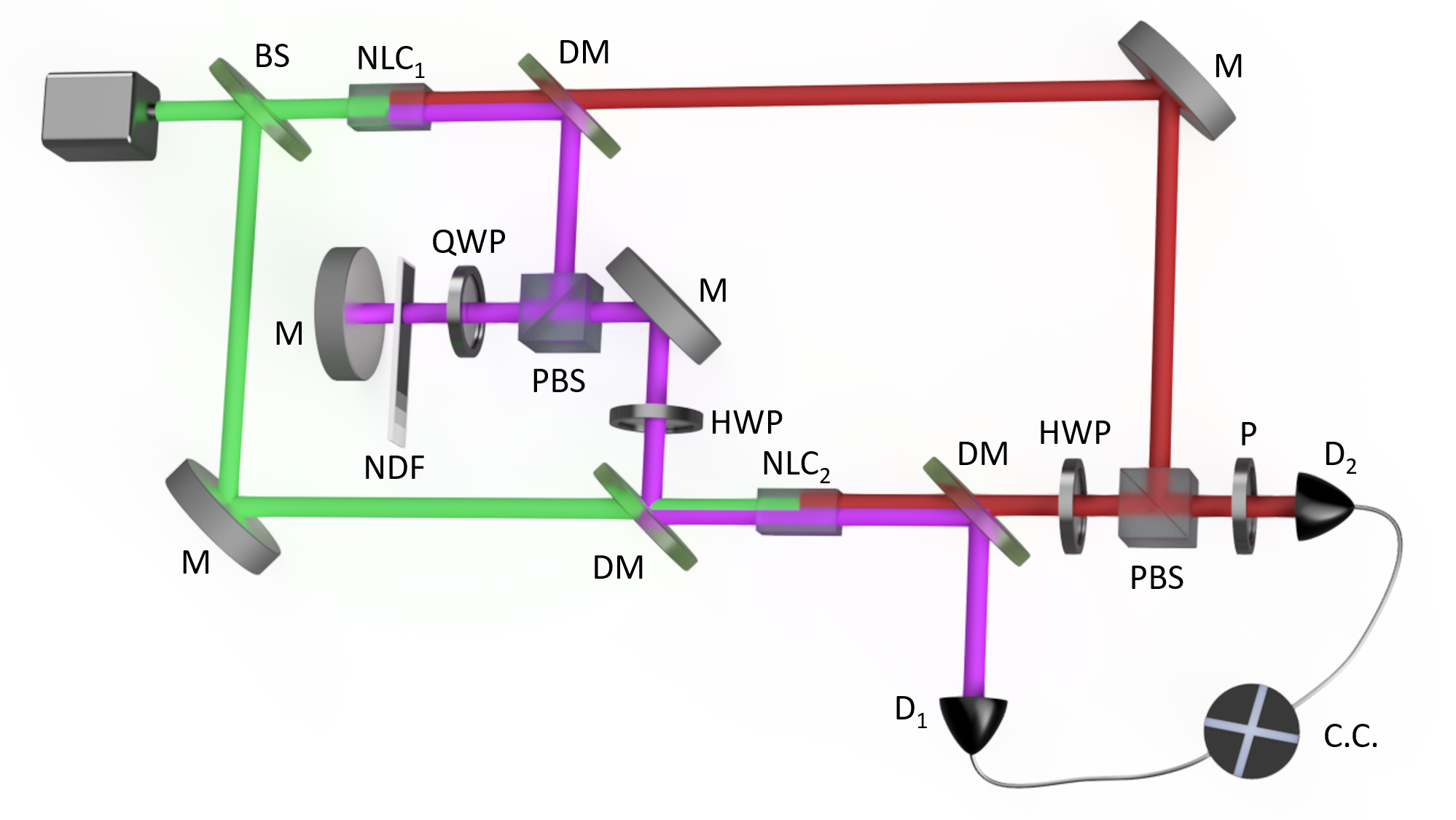}
 \caption{Experimental setup. For measuring the visibility $\mathcal{V}$ we use only detector D$_{2}$. For measuring second-order correlations, we detect coincidence counts (C.C.) between D$_{1}$ and D$_{2}$. BS: beam splitter; M: mirror; NLC$_{1,2}$: nonlinear crystals; PBS: polarizing beam splitter; DM: dichroic mirror; P: polarizer; QWP and HWP: quarter and half-wave plates; NDF: neutral density filter; $D_{1,2}$: single-photon detectors.} 
 \label{fig:path}
 \end{figure}
 
\section{Experimental setup}
It contains two identical sources, spatially separated, pumped by a continuous-wave laser with central wavelength $\lambda_{p}=532$ nm (see Fig. 2). The nonlinear crystals NLC$_{1,2}$ are two 20-mm-long periodically-poled lithium niobate crystals mounted in ovens. The central wavelengths of the signal and the idler photons are $\lambda_{s}=810$ nm and $\lambda_{i}=1550$ nm, respectively.

Idler $i_1$ experiences losses introduced with the help of a variable neutral density filter, which allows for varying the transmission coefficient $t$ before being injected into NLC$_2$. The two spatially overlapping idler photons, generated in both crystals and ideally indistinguishable, are coupled to a single-mode fiber and measured by detector $D_{1}$. The signal photons $s_1$ and $s_2$, with orthogonal polarizations, are recombined in the last polarizing beam splitter, fiber coupled, and measured by the detector $D_{2}$.

The erasure of distinguishability between paired photons generated in different nonlinear crystals is in general a highly demanding experimental task, and this poses challenges when striving to achieve high visibility values. Our group conducted an experiment using a similar configuration \cite{adam2018}, and we recorded a maximum visibility value close to $\mathcal{V} = 90\%$. In the induced coherence configuration aimed at demonstrating imaging with undetected photons \cite{lemos2014}, a maximum visibility of $\mathcal{V} = 77\%$ was measured. In the original induced coherence paper by Zou et. al. \cite{mandel1991PRL} from 1991, they measured a maximum visibility of $\mathcal{V} = 30\%$. These values from previous experiments show how difficult it is to compensate all the different degrees of freedom involved in the system that can provide unwanted path distinguishability.

In order to take into account the possible experimental distinguishability of idler modes coming from the first or second nonlinear crystals and propagating through path $i_3$, we write the quantum state of signal-idler photons as 
\begin{eqnarray}
    & & |\Psi \rangle_{si}=\frac{1}{\sqrt{2}}\,\ket{1}_{s_1} \ket{0}_{s_2} \Big[ r \ket{1}_{i_2}\ket{0}_{v_3} \ket{0}_{w_3} \nonumber \\
    & & + t \gamma \ket{0}_{i_2}\ket{1}_{v_3}\ket{0}_{w_3}+ t\sqrt{1-|\gamma|^2}  \ket{0}_{i_2}\ket{0}_{v_3}\ket{1}_{w_3} \Big]  \nonumber \\ 
    & & +\frac{1}{\sqrt{2}}\,\ket{0}_{s_1}\ket{1}_{s_2}  \ket{0}_{i_2} \ket{1}_{v_3} \ket{0}_{w_3},
\end{eqnarray}
where $v_3$ describes the spatio-temporal mode of the idler photon generated in NLC$_2$, and $w_3$ describes an orthogonal mode to $v_3$. $\gamma$ is the mode overlap between the spatio-temporal modes of idler photons generated in NLC$_1$ and NLC$_2$. The visibility of the interference of signal photons $s_1$ and $s_2$ is
\begin{equation}
\mathcal{V}=|\gamma||t|,
\label{newV}
\end{equation}
and for perfect overlap ($|\gamma|=1$) we recover the ideal value of $\mathcal{V}=|t|$.

 \begin{figure}[t!]
     \centering
     \includegraphics[width=\linewidth]{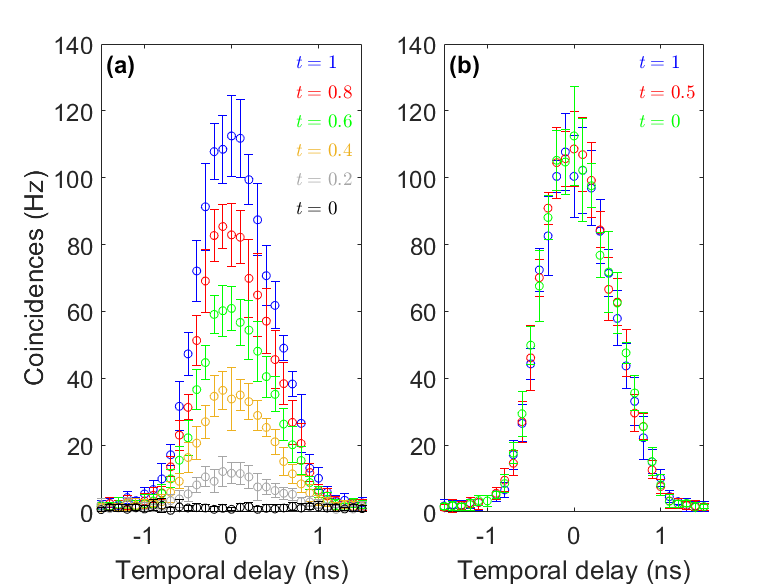}
     \caption[Experimental coincidence measurements]{Measured coincidence counts, $R_{13}$ and $R_{23}$, for different values of the transmission coefficient $t$ of the neutral density filter, as a function of the temporal delay $\tau$ between the photons. (a) Coincidence counts $R_{13}$ between signal $s_{1}$ and idler $i_3$ photons; (b) Coincidence counts $R_{23}$ between signal $s_{2}$ and idler $i_3$ photons.}
 \label{fig:coincidences}
 \end{figure}

The quantum states that describe the idler photon if the signal-idler pair is generated in the first or the second nonlinear crystals are
\begin{eqnarray}
    & &|\Psi_1 \rangle=r\ket{1}_{i_2}\ket{0}_{v_3} \ket{0}_{w_3}  + \gamma t \ket{0}_{i_2}\ket{1}_{v_3} \ket{0}_{w_3} \nonumber \\
    & &+t\,\sqrt{1-|\gamma|^2} \ket{0}_{i_2} \ket{0}_{v_3} \ket{1}_{w_3},
\end{eqnarray}
and
\begin{equation}
    |\Psi_{2} \rangle= \ket{0}_{i_2}\ket{1}_{v_3} \ket{0}_{w_3}.
\end{equation}
The distinguishability $\mathcal{D}$ between the two events is
\begin{equation}
    \mathcal{D}=\sqrt{1-|\gamma|^2\,|t|^{2}},
    \label{newD}
\end{equation}
and by substituting Eqs. (\ref{g13low}) and (\ref{g23low}) into Eq. (\ref{newD}) we obtain
\begin{equation}
    \mathcal{D}=\sqrt{1-|\gamma|^2\frac{g_{13}^{(2)}-1}{g_{23}^{(2)}-1}}.
\label{eq:distinguishability2}
\end{equation}
This is the expression that we need to consider in the experiment. For perfect overlap ($|\gamma|=1$) we recover Eq. (\ref{eq:distinguishability}). The goal is to evaluate $g^{(2)}_{13}$ and $g^{(2)}_{23}$ by measuring coincidence counts between signal $s_{1}$ and idler $i_3$, and between signal $s_{2}$ and idler $i_3$, as a function of $t$. The coincidence measurements are done between photons at different wavelengths. The signal photon is centered at 810 nm and the idler photon is centered at 1550 nm. See Appendix \ref{sec:app1} for a detailed explanation on how we measure coincidences. The parameter $|\gamma|$ will be determined experimentally through the measurement of the maximum visibility for $|t|=1$, which from Eq.~(\ref{newV}) yields $\mathcal{V}_{max}=|\gamma|=0.855$.

\section{Experimental results}
In an experimental setup, the rate of coincidence counts $R_{mn}$ in a coincidence detection time window $T_{R}$ is \cite{wolf1995}
\begin{eqnarray}
    & & R_{mn}= \int_t^{t+T_R} d\tau \langle\hat{a}_{m}^{\dagger}(t)\hat{a}_{n}^{\dagger}(t+\tau)\hat{a}_{n}(t+\tau)\hat{a}_{m}(t)\rangle= \nonumber \\
    & & =R_{m}R_{n}T_{R}\Big[ 1+\frac{T_{c}}{T_{R}}\,\Gamma_{mn}\Big],
\label{eq:R12}
\end{eqnarray}
where we make use of $g_{mn}^{(2)} = 1 + \Gamma_{mn}$. $R_{m}$ and $R_{n}$ are the measured single-photon flux rates, $T_c$ is the coherence time (inverse bandwidth) of signal-idler photons, and $\tau$ is the temporal delay between the photons. The coincidence detections are measured with $T_{R}=2.5$ ns for a total time of 30 s. The value measured in the experiment of the inverse bandwidth is $T_c=580$ fs. The measured single-photon flux rates are $R_{1,2} \sim 2000$ photons/s and $R_{3} \sim 2000$ photons/s.

 \begin{figure}[t!]
     \centering
     \includegraphics[width=\linewidth]{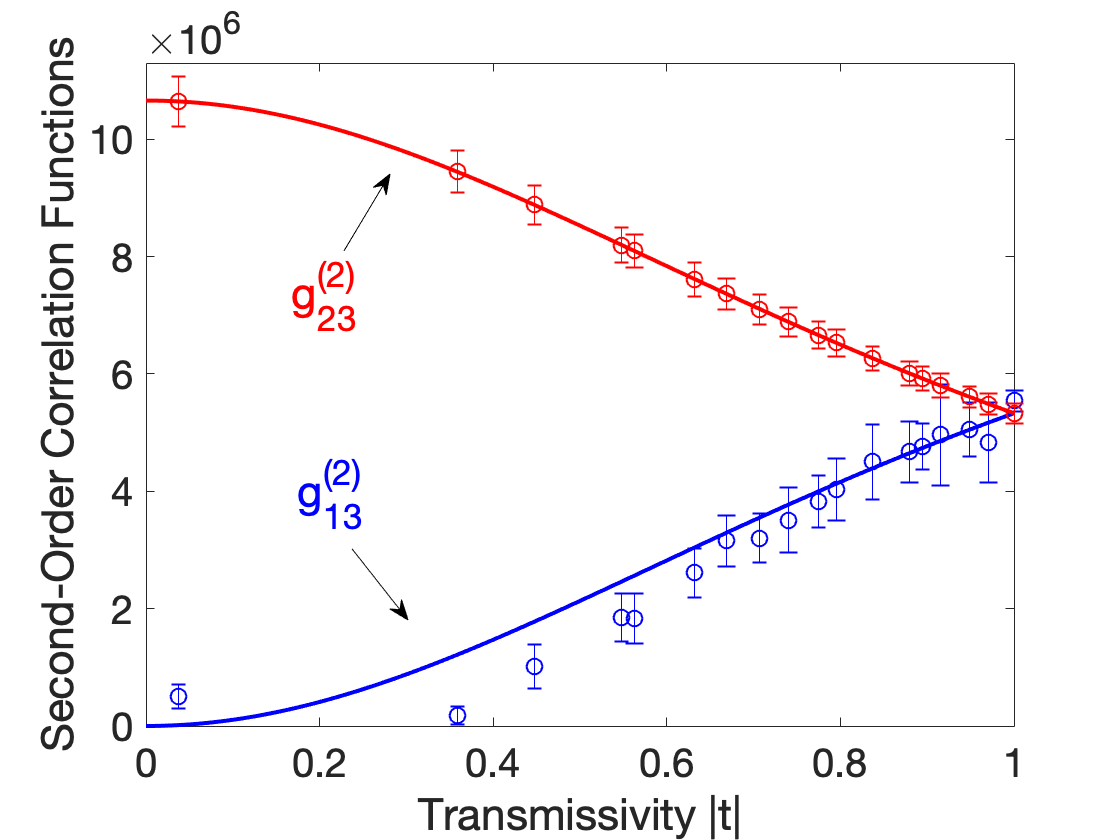}
     \caption[Experimental second-order correlation functions]{Normalized second-order correlation functions $g_{13}^{(2)}$ and $g_{23}^{(2)}$ as a function of the transmissivity $t$. Blue dots: experimental data for $g_{13}^{(2)}$. Red dots: experimental data for $g_{23}^{(2)}$. The theoretical curves are given by Eqs. (\ref{g13low}) and (\ref{g23low}). The error bars designate the standard deviation of the experimental measures.}
 \label{fig:correlations}
 \end{figure}

Figure \ref{fig:coincidences} shows the coincidence counts $R_{13}$ and $R_{23}$ measured as a function of the delay $\tau$ introduced between the signal and idler photons, for different values of the transmission coefficient $t$ introduced by the neutral density filter. In Fig. 3(a), we observe that the maximum of $R_{13}$ for $|t|=1$ is $R_{13}^{max}(|t|=1)=112.5$ coincidences/s, while for $|t|=0$ is $R_{13}^{max}(|t|=0)=5$ coincidences/s. Considering Eq. (\ref{g13low}), we obtain that
\begin{equation}
    \frac{R_{13}^{max}(|t|=1)}{R_{13}^{max}(|t|=0)} \sim \frac{T_c}{T_R} \frac{1}{|V|^2},
\end{equation}
where we make use of $T_c/(T_R\,|V|^2) \gg 1$. If we now consider Eq. (\ref{g23low}), we obtain
\begin{equation}
    \frac{R_{23}^{max}(|t|=1)}{R_{23}^{max}(|t|=0)} \sim 1,
\end{equation}
which is observed in Fig. 3(b): $R_{23}$ does not depend on the transmission coefficient $t$.

 \begin{figure}
     \centering
     \includegraphics[width=\linewidth]{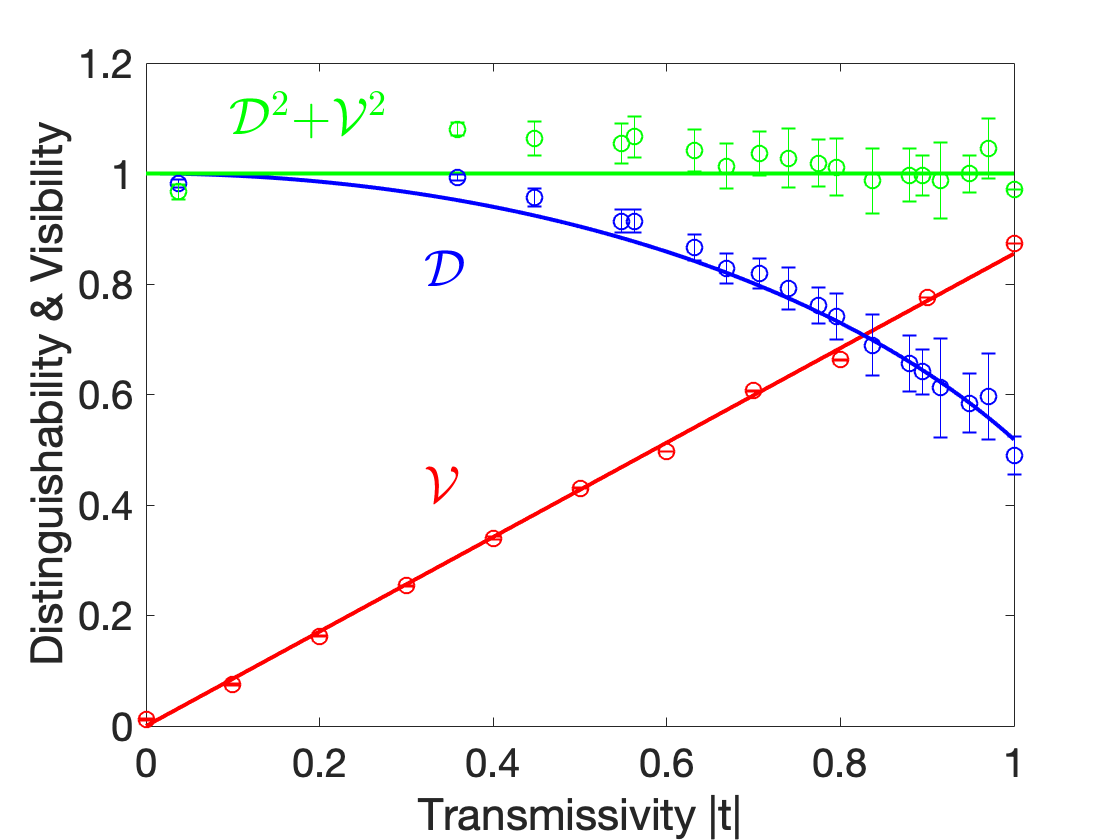}
     \caption[Experimental distinguishability and visibility results]{Experimental and theoretical relationship between the visibility $\mathcal{V}$ (red) and the distinguishability parameter $\mathcal{D}$ (blue) as a function of the transmissivity $t$. The circles represent experimental data and the solid lines represent theoretical predictions. The theoretical value of the visibility is given by Eq. (\ref{newV}), and the fit to the experimental values allow us to determine the value of the mode overlap parameter $\gamma=0.855$. The theoretical curve of the distinguishability is given by Eq. (\ref{newD}). The green circles represent the experimental values of $\mathcal{D}^2+\mathcal{V}^2$, and the solid green line is the theoretical prediction $\mathcal{D}^2+\mathcal{V}^2=1$. The experimental visibility values are recovered from the results demonstrated in Ref. \cite{adam2018}.}
 \label{fig:englert_exp}
 \end{figure}

We estimate the value of the variables $\Gamma_{13}$ and $\Gamma_{23}$ from the maximum of $R_{13}$ and $R_{23}$ in Fig. \ref{fig:coincidences}. The measured second-order correlation functions $g_{13}^{(2)}$ and $g_{23}^{(2)}$ are shown in Fig. \ref{fig:correlations}. The experimental results are in good agreement with the theoretical predictions shown in Eqs. \eqref{g13low} and \eqref{g23low}. 

Figure \ref{fig:englert_exp} shows the theoretical and experimental values of visibility $\mathcal{V}$ (red) and distinguishability $\mathcal{D}$ (blue) as a function of the transmission coefficient $t$. We also plot the values of the complementarity relationship $\mathcal{D}^2+\mathcal{V}^2$ (green). Visibility values are taken from the experiments described in \cite{adam2018}. The curve of experimental visibility allows us to determine the value of the parameter $\gamma$ in Eq. \eqref{newD}. We achieved a maximum visibility of $\mathcal{V}_{max} = 86\%$.

\section{Derivation of a complementarity relationship for all parametric gain regimes}
For the sake of simplicity, let us consider the distinguishability $\mathcal{D}$ given by Eq. (\ref{eq:distinguishability}). It was derived by comparing the parameter $\mathcal{D}$ obtained for the low parametric gain using the Schr\"odinger picture, i.e., $\mathcal{D}=\sqrt{1-|t|^2}$, with Eqs. (\ref{g13low}) and (\ref{g23low}), the second-order correlation functions obtained using the Heisenberg picture. The experimental demonstration of the complementarity relationship between $\mathcal{D}$ and $\mathcal{V}$ was done in the low parametric gain regime. 

Let us assume that Eq. (\ref{eq:distinguishability}) is still valid beyond the low parametric gain regime, where it was derived. Using the expressions of the normalized second-order correlation functions given by Eqs. \eqref{eq:g13} \& \eqref{eq:g23}, the distinguishability $\mathcal{D}$ [Eq. \eqref{eq:distinguishability}] for any gain regime is  (see Appendix \ref{sec:app3} for further details)
\begin{equation}
    \mathcal{D}= \sqrt{\frac{1-|t|^2+|t|^2(1-|t|^2)|V|^{2}}{1+2|t|^{2}|V|^{2}+|t|^{4}|V|^{4}}}.
\label{dist2}
\end{equation}
By using the relationship $|U|^2=1+|V|^2$, this equation is written as a function of the parameter $V$ only. For $|t|=0,1$ we have $\mathcal{D}=1$ and 0, respectively. 

The visibility ($\mathcal{V}$) of the interference pattern is not a good measure of coherence in the high parametric gain regime, since the flux rates of the signals $s_1$ and $s_2$ are different. We can see that for $t=1$, the visibility is $\mathcal{V}=1$ only in the low parametric gain regime.  Instead, we use the first-order degree of coherence $g_{12}^{(1)}$ between $s_1$ and $s_2$. See the Appendix \ref{sec:app2} for a derivation of the relationship between $\mathcal{V}$ and $g_{12}^{(1)}$. The degree of first-order coherence for any parametric gain regime is \cite{belinsky1992,wiseman2000}
\begin{equation}
g_{12}^{(1)}=|t|\sqrt{\frac{1+|V|^{2}}{1+|V|^{2}|t|^{2}}}. 
\label{g12}
\end{equation}
Making use of Eqs. (\ref{dist2}) and (\ref{g12}), both valid for all parametric gain regimes, it turns out that the equality  
\begin{equation}
    \mathcal{D}^{2}+[g_{12}^{(1)}]^{2}=1
\end{equation}
is fulfilled. This is the second important result of this paper. It constitutes a complementarity relationship between distinguishability and first-order coherence applicable beyond the single-photon regime, valid for all parametric gain regimes. 

\section{Conclusions} 
Most previously derived complementarity relationships are valid in the single-photon regime. In contrast to these prior studies, we have derived a relationship that extends beyond this regime, remaining valid for both low and high photon flux rates. We make use of a distinguishability parameter $\mathcal{D}$ based on second-order correlation functions, $g_{13}^{(2)}$ and $g_{23}^{(2)}$, that allows us to unveil a profound link between the first-order coherence of two waves (signals $s_1$ and $s_2$), i.e., their capacity to interfere, and the nature of correlations they exhibit with a third wave (idler beam). 

We have analyzed the induced coherence effect following the Schr\"odinger and Heisenberg pictures.  The juxtaposition of outcomes from both approaches enabled us to identify a potential distinguishability parameter. Our paper serves as an illustration of a discovery made through the combined utilization of the Schr\"odinger and Heisenberg pictures of Quantum Mechanics.

The results of this paper contribute to the ongoing discussion regarding the non-classicality of induced coherence \cite{zeilinger2019}. In the high parametric gain regime, the usual explanation centers on the induced phase coherence caused by the parametric amplification of the idler wave in the second nonlinear crystal \cite{shapiro2015}. The debate revolves around the fundamental explanation of induced coherence in the low parametric gain regime. On the one hand, the explanation is induced phase coherence, as in the high parametric gain regime. On the other hand, the alternative explanation is path indistinguishability of the interfering signal waves. In this paper, we introduce a complementarity relationship that is valid in all parametric gain regimes, and experimentally demonstrate its validity in the low-gain regime. This result shows that the concept of distinguishability, traditionally thought to be applicable solely in the single photon regime, can also be extended to the high parametric gain regime. Thus, this paper paves the way for new configuration proposals capable of experimentally demonstrating its validity in the high gain regime.

\begin{acknowledgments}
This work is part of the Research and Development Project No. CEX2019-000910-S, funded by MCIN/AEI/10.13039/ 501100011033/. It is supported by Fundaci\'o Cellex, Fundaci\'o Mir-Puig, and Generalitat de Catalunya through the CERCA program. We acknowledge financial support from project QUISPAMOL (Grant No. PID2020-112670GB-I00) funded by MCIN/AEI/10.13039/501100011033. This work is part of the project 20FUN02 “POLight”, which has received funding from the EMPIR program co-financed by the Participating States and from the European Union's Horizon 2020 research and innovation program. GJM was supported by the Secretaria d’Universitats i Recerca del Departament d’Empresa i Coneixement de la Generalitat de Catalunya and the European Social Fund (ESF) FEDER.

\end{acknowledgments}

\appendix

\section{Measurement of the second-order correlation functions}
\label{sec:app1}
The signal photons are detected by a silicon-based single-photon counting module SPCM-AQRH-14-FC (Perkin-Elmer), which detects single photons over the wavelength range of 400 to 1060 nm. The photon detection efficiency at 810 nm is approximately 60 $\%$. The signal photons are coupled to the detector via a single-mode fiber. Each photon detection generates a 30 ns width transistor-transistor logic (TTL) level electronic signal that is available at the output of a BNC connector.   

The idler photons are detected by a single-photon detection module id201 (idQuantique) based on indium gallium arsenide. This module detects IR photons with an efficiency of up to 25 $\%$. The idler photons are coupled via a single-mode fiber and each detection generates a TTL-type electronic signal of 100$\pm$10 ns width. It has an adjustable detection pulse width from 2.5 to 100 ns and a tunable delay between 0 and 25 ns.

The coincidence measurements are performed in the following way: the Perkin-Elmer output detection is used as the input trigger for the idler photon detection. As a result, an idler count is directly a coincidence. To make that happen, the photons' arrival times at their detectors must be fine tuned. Not only that, but one must additionally account for the detectors' response times as well as the delay provided by the BNC connections that link them. 

\begin{figure}[h!]
    \centering
    \includegraphics[width=\linewidth]{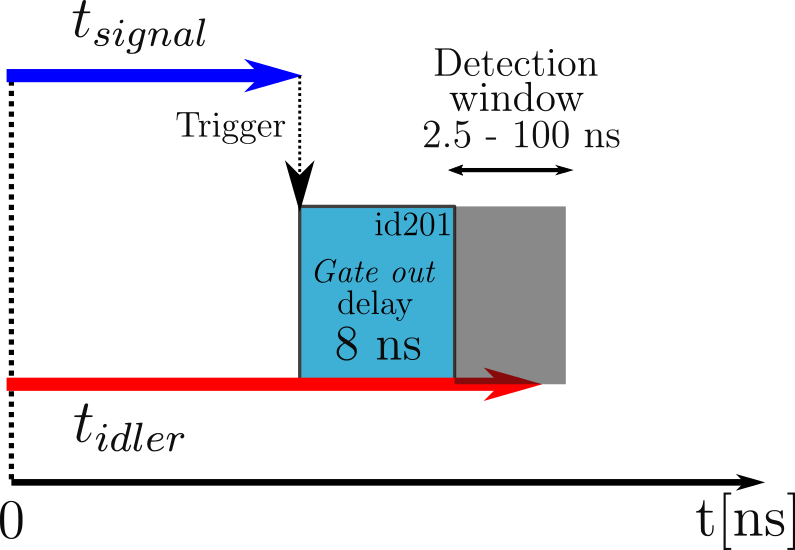}
    \caption{Temporal lines for both photons and the idQuantique id201 detector. The detection of the signal photon generates an electronic signal that serves as in input trigger to detect a coincidence with the id201 detector. To account for a coincidence, the time of arrival $t_{idler}$ of the idler photon to the detector needs to be adjusted so that it \textit{falls} within the detection window.}
    \label{fig:coincidences_times}
\end{figure}

Figure \ref{fig:coincidences_times} shows a sketch with the timelines of both photons and the id201 detector. The time of arrival of the idler photon to the detector needs to be adjusted so that the condition: $t_{idler} \geq t_{signal} + 8$ ns, is fulfilled. To do this, we must use the right fiber and coaxial cable lengths. The signal and the idler photons' single-mode optical fiber lengths in the experiment are $l_{signal}=2$ m and $l_{idler}=12$ m, respectively. The length of the BNC coaxial cable that connects both avalanche photodiodes is $l_{BNC}=0.40$ m.

\section{Relationship between the first-order correlation function and the visibility of an interference pattern}
\label{sec:app2}
Consider an optical signal, with associated quantum operator $\hat{a}_3$, that is the superposition of two signals with associated operators $\hat{a}_1$ and $\hat{a}_2$:
\begin{equation}
    \hat{a}_3=\hat{a}_1 \exp(i \varphi_1)+\hat{a}_2 \exp(i \varphi_2),
\end{equation}
where $\varphi_{1,2}$ are the phases acquired by signals $1$ and $2$ during propagation. The average number of photons $N_{3}$ is
\begin{eqnarray}
    & & N_3\equiv \langle \hat{a}_3^{\dagger} \hat{a}_3 \rangle =  \\
    & & = N_1 +N_2+2 N_1^{1/2}\, N_2^{1/2}\, |g_{12}^{(1)}|\, \cos (\varphi_2-\varphi_1+\varphi_g), \nonumber
\end{eqnarray}
where $N_{1,2} \equiv \langle \hat{a}^{\dagger}_{1,2} \hat{a}_{1,2} \rangle$ and we write the first-order correlation function $g_{12}^{(1)} \equiv \langle \hat{a}_1^{\dagger} \hat{a}_2 \rangle/[N_1^{1/2}\, N_2^{1/2}]$ as $g_{12}^{(1)}=|g_{12}^{(1)}| \exp(i\varphi_g)$.

As a function of the phase difference $\Delta \varphi=\varphi_2-\varphi_1$, the value of $N_3$ oscillates between the maximum
\begin{equation}
N_3^{max}=N_1 +N_2+2 N_1^{1/2}\, N_2^{1/2}\, |g_{12}^{(1)}|
\end{equation}
and the minimum
\begin{equation}
N_3^{min}=N_1 +N_2-2 N_1^{1/2}\, N_2^{1/2}\, |g_{12}^{(1)}|,
\end{equation}
so the visibility of the interference fringes is
\begin{equation}
\mathcal{V}=\frac{N_3^{max}-N_3^{min}}{N_3^{max}+N_3^{min}}=\frac{2 N_1^{1/2}\, N_2^{1/2}\, }{N_1+N_2}\, |g_{12}^{(1)}|.
\end{equation}
We define the parameter $\Delta$, a measure of the energy unbalance between signals $1$ and $2$, as \cite{friberg2017, QVE17}
\begin{equation}
    \Delta=\frac{\big|N_1-N_2\big|}{N_1+N_2}.
\end{equation}
Taking into account that
\begin{equation}
   \frac{2N_1 N_2}{(N_1+N_2)^2}= 1-\Delta^2,
\end{equation}
we can write
\begin{equation}
\mathcal{V}= \sqrt{1-\Delta^2}\, |g_{12}^{(1)}|.
\label{D}
\end{equation}
For no energy unbalance, $N_1=N_2$, $\Delta=0$ and $\mathcal{V}=|g_{12}^{(1)}|$. The visibility is a direct measure of the value of the first-order correlation function. When $N_1 \ne N_2$,  the value of the first-order correlation function should be derived from the measurement of the visibility with the help of Eq. \eqref{D}.
\vspace{0.1em}

\section{Derivation of Eq. (23) of the main text}
\label{sec:app3}
For the sake of clarity, in this appendix we give an explicit derivation of Eq. (\ref{dist2}) of the main text:
\begin{equation}
    \mathcal{D}=\sqrt{\frac{1-|t|^2+|t|^{2}(1-|t|^{2})|V|^{2}}{1+2|t|^{2}|V|^{2}+|t|^{4}|V|^{4}}},
\label{C1}
\end{equation}
where  $\mathcal{D}$ is defined as
\begin{equation}
    \mathcal{D}=\sqrt{\frac{g_{23}^{(2)}-g_{13}^{(2)}}{g_{23}^{(2)}-1}}.  
\label{C2}
\end{equation}
The expressions of the second-order correlations functions $g_{13}^{(2)}$ and $g_{23}^{(2)}$ [Eqs.  \eqref{eq:g13} \&  \eqref{eq:g23} in the main text] are
\begin{equation}
 g_{13}^{(2)}=1+\frac{|t|^2\,|U|^4}{1+|t|^2 |U|^2}\, \frac{1}{|V|^2}
\label{C3}
\end{equation}
and
\begin{eqnarray}
    & & g_{23}^{(2)}=1+ \\
    & &+\frac{|t|^{4}|U|^{6}+2|t|^{2}|U|^{4}-2|t|^{4}|U|^{4}+|U|^{2}(1-|t|^{2})^{2}}{|V|^{2} \big[ 1+ |t|^{2}|V|^{2} \big]\big[ 1+|t|^{2}|U|^{2}\big]}.  \nonumber
\label{C4}
\end{eqnarray}
If the denominators of the expressions $g_{23}^{(2)}-g_{13}^{(2)}$ and $g_{23}^{(2)}-1$ in Eq. \eqref{C2} are made equal, and the relation $|U|^2=1+|V|^2$ is used, the numerator of $g_{23}^{(2)}-1$ can be written as:
\begin{eqnarray}
    & & g_{23}^{(2)}-1= \nonumber \\
    & & =|t|^{4}|U|^{6}+2|t|^{2}(1-|t|^2)\, |U|^{4}+\left(1-|t|^{2} \right)^{2}\,|U|^{2}= \nonumber \\
    & & = |t|^{4} \left( 1+3|V|^2+3|V|^4+|V^6| \right)+ 2|t|^{2}(1-|t|^2)    \nonumber \\
    & &  \times \left( 1+2|V|^{2}+|V|^4 \right)+\left(1-|t|^{2} \right)^{2}\,\left( 1+|V|^{2} \right)= \nonumber \\
    & & = \left( 1+2|t|^2|V|^2+|t|^4|V|^4\right)\ \left(1+|V|^2 \right),
\label{C5}
\end{eqnarray}
\vspace{0.1em}

and the numerator of $g_{23}^{(2)}-g_{13}^{(2)}$ is
\begin{eqnarray}
    & & g_{23}^{(2)}-g_{13}^{(2)}= \nonumber \\
    & & = \left( 1+2|t|^2|V|^2+|t|^4|V|^4\right)\ \left(1+|V|^2 \right) \nonumber \\
    & & - |t|^2\ \left(1+|t|^2|V|^2 \right)\ \left(1+|V|^2 \right)^2= \label{numerator} \\
    & & =\left[  1-|t|^2+|t|^2 (1-|t|^2)\, |V|^2 \right] \left(1+|V|^2 \right) \nonumber, 
\end{eqnarray}
Dividing Eq. \eqref{numerator} by Eq. \eqref{C5} leads directly to Eq. \eqref{C1}, which corresponds to Eq. \eqref{dist2} in the main text.

\nocite{*}

\providecommand{\noopsort}[1]{}\providecommand{\singleletter}[1]{#1}%

\end{document}